\documentclass[11pt]{article}
\usepackage{amssymb,amsmath,cite}
\def\m{\mu}
\def\n{\nu}
\def\e{\epsilon}
\def\a{\alpha}
\def\o{\omega}
\def\d{\delta}
\def\half{{1 \over 2}}
\def\pa{\partial}
\def\minv{m^{-1}}
\def\bfr{\mathbf{r}}
\catcode`\@=11
\def\m{\mu}
\def\n{\nu}
\def\e{\epsilon}
\def\a{\alpha}
\def\half{{1 \over 2}}
\def\pa{\partial}
\def\minv{m^{-1}}
\@addtoreset{equation}{section}
\def\theequation{\arabic{section}.\arabic{equation}}
     
\catcode`\@=11
\def\thesection{\arabic{section}}

\def\appendix{\setcounter{section}{0}
        \def\thesection{Appendix}
        \def\theequation{\Alph{section}.\arabic{equation}}}
\def\section{\@startsection{section}{1}{\z@}{3.5ex plus 1ex minus
   .2ex}{2.3ex plus .2ex}{\large\bf}}
   
\long\def\@makefntext#1{\parindent 0cm\noindent
\hbox to 1em{\hss$^{\@thefnmark}$}#1}

\newcommand{\captionfonts}{\small}
\makeatletter  % Allow the use of @ in command names
\long\def\@makecaption#1#2{%
  \vskip\abovecaptionskip
  \sbox\@tempboxa{{\captionfonts #1: #2}}%
  \ifdim \wd\@tempboxa >\hsize
    {\captionfonts #1: #2\par}
  \else
    \hbox to\hsize{\hfil\box\@tempboxa\hfil}%
  \fi
  \vskip\belowcaptionskip}
\makeatother   % Cancel the effect of \makeatletter 

\begin{document}
\begin{titlepage}
\vspace{.5in}
\begin{flushright}
September 2008\\
\end{flushright}
\vspace{.5in}
\begin{center}
{\Large\bf
Distended Topologically Massive Electrodynamics} 

\vspace*{.4in}
{S.~Deser}\footnote{\texttt { deser@brandeis.edu}}\\
       {\small\it Physics Department, Brandeis University, Waltham, MA 02454}\\
       {\small\it Lauritsen Laboratory, California Institute of Technology, Pasadena, CA 91125}\\
\vspace*{2ex}
      {Dedicated to the memory of Wolfgang Kummer, intrepid explorer of (a somewhat) lower dimension. To appear in Kummer memorial volume.} 
\end{center}

\vspace{.5in}
\begin{center}
{\large\bf Abstract}
\end{center}
\begin{center}
\begin{minipage}{4.5in}
{\small
We extend topologically massive electrodynamics, both by adding a higher derivative action to cast the entire three-term model in Chern-Simons (CS) form, and by embedding it in an AdS background. It can then be written as the sum of two CS terms, one of which vanishes at the ``chiral" point, in analogy with its gravitational topologically massive counterpart. Separately we treat pure CS electrodynamics plus Einstein gravity interacting with point sources. The gravity/vector field equations decouple; their solutions are the familiar exterior ``conical" metric and vector potentials. 
}
\end{minipage}
\end{center}
\end{titlepage}
\addtocounter{footnote}{-1}

\section{Introduction} \label{intro}

Theories involving Chern-Simons (CS) terms have remained popular ever since their introduction, in both gravitational and vector incarnations, over a quarter-century ago [1]. Most recently, there has been a great deal of work on extension of the original models from flat to AntideSitter (AdS) backgrounds [2,3] Separately, standard topologically massive electrodynamics (TME) has been augmented by higher derivative, but CS-like, terms [4]. Here we combine these two generalizations to express TME as a ``2-CS" chiral sum.

Our second topic is ``pure CS" - combined Einstein plus vector CS actions - in presence of point masses and charges. In both cases, the field equations exhibit a field-current identity. Further, the two field sectors decouple, so the total system remains soluble. Accordingly, the resulting metric and vector potential have conical structure.

 \section {E(xtended)TME }

Gravitational (tensor) and electrodynamical (vector) field models are often both similar and different; this is also true of their CS properties. The most relevant difference is that the gravitational CS actions is of third derivative order - higher than the Einstein action - whereas it is the opposite for vectors, whose CS term is first order, lower than the Maxwell action's.  This is reflected in constructing their ``pure CS" extensions, requiring respectively cosmological/higher derivative additions to the topologically massive two-term models. However, even for TME, as we shall see, adding a cosmological background simulates TMG in this context. For simplicity, we consider only abelian TME here. 

We begin with the vector CS action,
\begin{equation}
  \label{eq:1}
I_{CS}(B) = \int d^3 x \, \e^{\m \n \a} B_\m \pa_\n B_\a \, .
\end{equation}
The resulting field equation,
\begin{equation}
  \label{eq:2}
F^\m(B) \equiv {1 \over 2} \e^{\m \n \a} F_{\n \a}(B) = 0
\end{equation}
states that field space is ``flat," with a pure gauge vector potential. Next, generalize $B_\mu$ to be a combination
\begin{equation}
  \label{eq:3a}
B^\pm_\m(A) \equiv m^{-1/2} F_\m(A) \pm m^{1/2} A_\m
\end{equation}
of the fundamental variable $A_\mu$. The parameter $m$ has dimensions of mass, needed to give $A_{\mu}$  its canonical dimension. We have also allowed for separate combinations $B_{\pm}$ which could be further generalized by allowing for two separate mass values, $m _{\pm}$. Our conventions are $(-++)$ signature, $\epsilon^{012} = 1$; the background is (initially) flat.

The action \eqref{eq:1} with $B(A)$ as in \eqref{eq:3a} consists of three terms,
\begin{align}
  \label{eq:4}
I_{CS}^\pm\left( B^\pm(A) \right) &= m^{-1} \int d^3 x \, \e^{\m \n \a} \left \{ F_\m (A) \pa_\n F_\a(A) + m^2 A_\m \pa_\n A_\a \right \} \pm  \int d^3 x \, F_{\m\n}^2(A) \nonumber \\
	&\equiv \{ m^{-1} I_{ECS} + m I_{CS} \} \mp 4 I_{MAX}\,.	
\end{align}
This result confirms for spin 1 that ``everything is CS" in $D=3$; even the Maxwell action is the difference of two CS terms. The two-mass generalization would permit even more flexibility in the relative coefficients.

The dynamics of the various three-term models in~\eqref{eq:4} was analyzed in [4], whose results we summarize for completeness. Pure $I_{ECS}$ leads to a null-propagating field strength, $\Box F_{\mu \nu} {=0}$, and hence does allow excitations, distinct from Maxwell's where $F^{\mu\nu}$ is of course also divergence-free. This model differs from pure $I_{CS}$ in not being topological: for example, its action is metric - dependent and more fundamentally, it shows no interesting large gauge behavior, owing to the pure field-strength dependence, even in the nonabelian version. The combination $I_{ECS}$ + $I_{MAX}$ also differs from the original TME: it contains a massive ghost excitation as well as the photon mode. Combining $I_{ECS}$ + $I_{CS}$ does not add further excitations to that of $I_{ECS}$ alone: instead, the field strength now propagates massively. Finally, the full three-term action depends on the two relative internal coefficients, and there are in general three masses, though there can be a degeneracy for suitable tuning. In all cases a ghost is unavoidable. 

The above results are easily checked explicitly from the field equations, with the usual decompositions of the potentials into invariant, pure gauge and constraint components: since each term is separately gauge invariant, all excitations are as well, and depend only on the transverse vector potential, effectively the indexless scalar in $ A^{T}_{i} = \epsilon^{ij} \partial_{j} S$.

\section{C(osmological)ETME}

We now introduce a nontrivial - AdS - gravitational background. The relevant aspect of this generalization is the appearance of a second dimensional parameter, the cosmological constant, ${\Lambda} \equiv - {\ell}^{-2}$.

Let us modify our previous discussion to follow the gravitational ``2-CS" formulation of [2]\footnote {This was also found by D. Grumiller and R. Jackiw (unpublished.)}. There, the variable corresponding to $B^{\pm}$ of \eqref{eq:3a} is a very similar combination of $B_{\mu}^{\pm}$, namely $\omega_{\mu}^{ab}(e) \pm \ell^{-1} \epsilon^{abc} e_{\mu c}$, 
where  ${\omega(e)}$ is the spin connection constructed from the dreibein $e_{\mu c}$. Note the required inverse length, which we mostly set to unity. In this fashion, we get two gravitational CS combinations
\begin{equation}
 \label{eq:5}
 I_\pm [ \o (e)\pm e ] = I_{GCS}[ \o (e) ] \mp I_{GR} [ e ],
\end{equation}
where $ I_{GCS} \sim \half \int d^3 x ( \e \o \pa \o + \ldots ) $ , is the (third derivative) gravitational CS term, $I_{GR}$ is the Einstein action including the cosmological term (proportional to $\Lambda$) but with the ``wrong" sign required by TMG to ensure ghost freedom. To construct the cosmological topologically massive action, a mass parameter ${m}$---distinct from $\ell^{-1}$---is introduced by hand to yield, from~\eqref{eq:5},
\begin{align}
  \label{eq:6}
  2 I_{CTMG} &= (1+\minv)I^- + (1 - \minv)I^+ \nonumber \\
  &= - \int d^3 x \, \sqrt{-g} (R- \Lambda) + \minv I_{GCS}
\end{align}
in Planck units. This CS doublet degenerates to a single term at either ``chiral" value $m=\pm1$. 

Returning to our vector case, we define the extended variable to be the $m=\ell^{-1}$ value of \eqref{eq:3a},
\begin{equation}
  \label{eq:3b}
B^\pm_\m(A) = f_\m(A) \pm A_\m \,.
\end{equation}
 [The other effect of the nontrivial background, say $g_{\mu \nu} = \phi^{2} \eta_{\mu \nu}$, is that $f_{\mu}(A)$ is here a covariant vector, like $A_{\mu}$ so it acquires a factor $\phi^{-1}$ . Hence  $I_{ECS}$ is scaled by $\phi^{-2}$, while $I_{MAX}\sim \phi^{-1}$ and of course $I_{CS}$ is metric-independent. These extra factors are not directly relevant to our discussion.] The analog of \eqref{eq:4} is simply obtained by replacing $m$ by $\ell^{-1}$  there.  Consequently, the CETME action is the combination
\begin{equation}
  \label{eq:7}
  8 I = (4 m \ell - 1)I^+ + (4 m \ell + 1)I^-
\end{equation}
where $m$ is the mass parameter of TME, and we have restored $\ell$ explicitly. This parallels the gravitational form \eqref{eq:6} except for the dimensionally dictated 
$m\rightarrow {\frac {1}{m}}$
there. This is the 3-term analog of TMG, and all three terms must be present. The $m \rightarrow {0}$ limit is of course Maxwell, but ordinary 2-term TME is obtained only in the singular $\ell \rightarrow {0}$ limit, while for gravity, it is the infinite mass limit that yields the (cosmological) Einstein action.    

At the chiral points $4m  \ell = \pm{1}$, one of the actions vanishes, exactly as for chiral gravity. The physics of ordinary two-term CTME at the chiral point is laid out in [2], where it is shown to be in one-to-one correspondence with linearized CTMG at the latter's chiral point.

\section {Sources}

So far, we have studied our models in a source-free context. We now include sources, in a particular, ``pure CS", gravity plus CS context. 

It is instructive to first analyze the relevant similarities to - and differences from - the gravitational case.  Recall that for spin two, the highest, third derivative, term is the gravitational CS action; its variation is the Cotton - conformal curvature - tensor, whose vanishing implies the metric is conformally flat. The Einstein action instead, effectively resembles that of pure vector CS: in both cases, their variations are the respective ``curvature terms," whose vanishing implies field flatness.  [The Maxwell term has no gravitational analog since it does describe a single physical excitation.] It is therefore really the Einstein and vector CS terms that corresponded most closely in the two systems. In each case, there is a field - current identity, respectively
\begin{equation}
  \label{eq:8}
  G_{\m \n}=T_{\m \n}
\end{equation}
\begin{equation}
  \label{eq:9}
  F^\m = j^\m
\end{equation}
where the Einstein tensor in \eqref{eq:8} equivalent (being its double dual) to the full curvature, so that spacetime is flat away from sources, and there is no interaction among localized masses [5]. Similar considerations hold in presence of a $\Lambda$ term, except that the exterior now has constant curvature [6]. The same holds for the field strength in \eqref{eq:9}, and non-interaction among charges. Note the counterintuitive property of \eqref{eq:9} that charges create magnetic, while currents create electric, fields: $F^{0}$ is the magnetic field 
$\epsilon^{ij} F_{ij}$, while $F^{i}$ is the electric field $\epsilon^{ij} E_{j}$. Point charges are represented by a current
\begin{equation}
  \label{eq:10}
  j^\m = \Sigma e_A u^\m_A(t) \d^2 (\bfr_\m -\bfr_A(t)) \,.
\end{equation}
Note that $j^\m$ is actually a metric-independent contravariant vector density just like $F^{\mu}$, so the tensor and vector equations are totally independent. Current conservation alone requires the particle worldlines to be continuous (albeit not necessarily future timelike), while covariant conservation constrains any point-like stress tensor to be that of (a sum of) particles [7]. As in gravity, while there is no interaction, the large-scale ``geometry" is affected by the configurations: in gravity these are the well-known metrics with conical singularities at the sources, together with their boosted generalizations, as discussed in [5], and similarly, as we now see, for the vector potentials \footnote {A more detailed perspective on CS electrodynamics with point charges may be found in [8].} 

The simplest case is a single static charge at the origin,
\begin{equation}
  \label{eq:11}
  j^0 = e \d^2(\mathbf{r})
\end{equation}
which generates a pure magnetic field, $\epsilon^{ij} F_{ij} \sim e \d^{2}(\bfr)$ whose vector potential is
\begin{equation}
  \label{eq:12}
  A_i = {\frac {-e} {2 \pi}} \e^{i j} \pa_j \ln r + \pa_i \a \,.
\end{equation}
Clearly, the potential is a superposition of such contributions if there are more static particles. Note that there is no self-force here, since the $\int A_{\mu} j^{\mu}$ term vanishes identically. The configuration can be sampled through its Aharonov-Bohm phase, proportional to the sum of the charges. A moving particle will generate an electric field as well; for a single source,
\begin{equation}
  \label{eq:13}
  F^i = \e^{i j} E_j = \e^{i j} \dot{A}_j = e u^i \d^2 (\mathbf{r} - \mathbf{r}(t)).
\end{equation}
This corresponds to a time-dependent vector potential $A_{i}(t)$ in $A_{0} = {0}$ gauge, with step-function behavior. 

The Einstein + CS + particle system is now easy to solve; as noted, the vector CS term, being topological, is metric-independent, as is the particle's $j^{\mu}$, so the combined field equations decouple,
\begin{equation}
  \label{eq:14}
  G^{\m \n} = m u^\m u^\n \d^2 (\bfr), \qquad F^\m = e u^\m \d^2(\bfr)
\end{equation}
and reduce, for the single static charge, with $u^{\mu} = \delta^{0 \mu}$, to the usual conical space with deficit angle proportional to the source's mass, but independent of any charge properties, and a ``conical" vector potential proportional to the total charge but independent of mass, as described by \eqref{eq:12}. The extension to superposition of several static particles is immediate, though there are interesting global geometric complications and limitations on the mass -  and perhaps also (color) charge - parameters, and even more for moving particles, despite the absence of true dynamics. Irrespective of the details of generic, distributed, interior sources, the exterior fields are those of a single particle with total mass and charge.

\section {Summary}

We have discussed two separate problems: the primary one was to obtain a ``pure CS" formulation of vector models in $D=3$ to include the Maxwell action. This required addition of a third-derivative CS-like term. In an AdS  background, the same procedure further allowed for a two-CS formulation using the freedom afforded by presence of two mass parameters ($m,\ell^{-1}$). Here as in gravity, at either special ``chiral" point, one of the two CS terms vanishes. This is also the common point for which TMG and TME equations can be put into one-one correspondence.

Our second topic was that of ``pure-CS" in the literal sense of keeping only the CS vector term along with its corresponding gravitational term, the Einstein action (with or without $\Lambda$). This two-field system was coupled to charged point masses. Because the two fields are entirely decoupled (CS being topological), the resulting configurations are separate conical metric and vector potential ``spaces", with (known) interesting geometric complications in the gravitational sector. The nonabelian vector side should also prove of interest.

\vspace{4ex}
I thank my collaborators  on [2], S. Carlip, A. Waldron, and D. Wise, whose insights there have also been useful here. This work was supported by NSF grant PHY 07-57190 and DOE grant DE-FG02-92ER40701.

\newpage

\begin{center}
{\bf REFERENCES}
\end{center}
\noindent 1. S.\ Deser, R.\ Jackiw, and S.Templeton, {\it Ann.Phys.}, {\bf 140}, 372, 1982; {\it Phys.Rev.Lett.}, {\bf 48}, 975, 1982.\\
2. S.\ Carlip, S.\ Deser, A.\ Waldron, and D.\ Wise, hep-th 0803.3998; 0807.0486, {\it Phys.Lett.}, {\bf B666}, 272, 2008. \\
3. P.\ Kraus and F.\ Larsen, hep-th/0508218,{\it JHEP} {\bf 0601}, 22, 2006; 
S.N.\ Solodukhin, hep-th/0509148 {\it. Phys.Rev.}, {\bf D74}, 024015, 2006;
W.\ Li, W.\ Song, and A.\ Strominger, hep-th 0801.4566;
A.\ Strominger, hep-th 0808.0506;
D.\ Grumiller and N.\ Johansson, hep-th 0805.2610;
W.\ Li, W.\ Song, and A.\ Strominger, hep-th 0805.3101;
P.\ Baekler, E.W.\ Mielke, and F.W.\ Hehl,{\it Nuovo Cim.} {\bf B107}, 91, 1992;
S.\ Cacciatori, M.\ Caldarelli, A.\ Giacomini, D.\ Klemm, and D.S.\ Mansi, hep-th/0507200,{\it J. Geom. Phys.}, {\bf 56}, 2523, 2006;
S.\ Carlip, hep-th 0807.4152;
D.\ Grumiller, R.\ Jackiw, and N.\ Johansson, hep-th 0806.4185, this Volume;    
G.\ Giribet, M.\ Kleban, and M.\ Porrati, hep-th 0807.4703v2.\\
4. S.\ Deser and R.\ Jackiw, hep-th/9901125, {\it Phys. Lett.} {\bf B451}, 73, 1999.\\
5. S.\ Deser, R.\ Jackiw, and G.\ t'Hooft, {\it Ann. Phys.}, {\bf 152}, 220, 1984.\\
6. S.\ Deser and R.\ Jackiw, {\it Ann. Phys.}, {\bf 153}, 405, 1984\\
7. W. Tulczyew, {\it Acta. Phys Polon.} {\bf 18}, 393, 1971;
C.\ Aragone and S.\ Deser, {\it Nucl. Phys.}, {\bf B92}, 327, 1975;
M.\ Gurses and F. Gursey, {\it Phys. Rev.}, {\bf D11}, 967, 1975.\\
8. R.\ Jackiw, {\it Ann. Phys.}, {\bf 201}, 83, 1990.

\end{document}